# Landau–Bogolubov Energy Spectrum of Superconductors

**L.N. Tsintsadze[1] and N.L. Tsintsadze[1,2]**

1. Department of Plasma Physics, E. Andronikashvili Institute of Physics, Tbilisi 0128, Georgia
2. Faculty of Exact and Natural Sciences, Tbilisi State University, Tbilisi 0105, Georgia

**Abstract**
We demonstrate that a dispersion relation of elementary excitations in the Fermi liquid as a superconductor is identical to the one in a quantum liquid $He\ \underline{II}$. Hence, we show that the superconductivity is, in fact, the same as superfluidity, but for charged particles.

It is well known that at temperatures $1 - 2\ K°$ only two quantum neutral liquids exist in nature, the isotopes of helium $^4He$ the Bose liquid and $^3He$ the Fermi liquid. The peculiarly weak interaction between the helium atoms is reason for helium to remain liquid. Landau has formulated the theory of Bose and Fermi liquids in the remote past [1-3]. The theory of Bose liquid $^4He$ was created by Landau [1] following Kapitza's [4] discovery of the superfluidity of liquid helium $^4He$. Bogolubov derived the energy spectrum for the elementary excitations in a quantum Bose liquid, and formulated the microscopic theory of the superfluidity of liquid helium $^4He$ in Ref. [5]. The same year Landau [2], in order to explain Peshkov's experiment [6], empirically constructed a dispersion relation of elementary excitations in a quantum liquid $He\ \underline{II}$, which is illustrated in Fig. 1

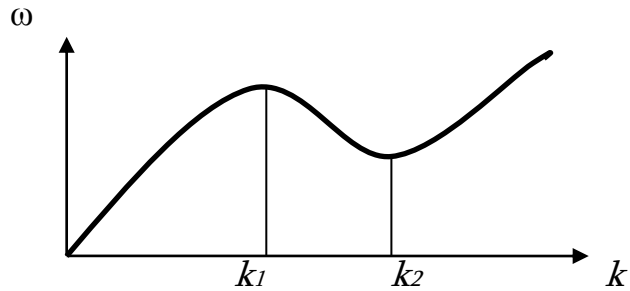

Fig. 1

Here $\omega$ and $k$ are the frequency and value of the wave vector, respectively.

The Fig.1 exhibits that after a linear increase (phonon part) the $\omega(k)$ reaches a maximum at a certain value $k_1$, then decreases and passes through a minimum at a certain value $k_2$ (roton part). The qualitative analyses of such spectra were given by Feynman in Ref. [7]. Later Feynman and Cohen [8] carried out a theoretical calculation of the spectrum of the elementary excitations.

Writing the function $\omega(k)$ near the minimum $k_2$, $\omega(k)$ can be expanded in powers $(k - k_2)$, as

$$\omega(k) = \omega(k_2) + \frac{1}{2}\left(\frac{\partial^2 \omega}{\partial k^2}\right)_{k=k_2} (k - k_2)^2, \text{ or } \varepsilon(k) = \hbar\omega(k) = \Delta_0 + \frac{\hbar^2(k-k_2)^2}{2m^*}, \qquad (1)$$

where $\hbar$ is the Plank constant divided by $2\pi$, $\Delta_0 = \hbar\omega(k_2)$ is the energy gap and $m^*$ is the effective mass of the quasi-particle.

These types of quasi-particles are called rotons. It should be emphasized that both phonons and rotons are quasi-particles constituting different parts of the same curve, and there is a continuous transition from one to the other.

Above we mentioned $^4He$ and $^3He$ as the only real neutral quantum liquids found in nature. However, there is also the quantum electron liquid, for instance, the conduction electrons in metals, degenerate semiconductors and semimetals.

For real Fermion liquids the particle interaction and the exclusion principle act simultaneously. Note that a superconducting electron gas is drastically modified by the particle interactions.

In this Letter, we show that in a superconductor, a non-ideal Fermi gas with attraction between electrons, the dispersion relation of the elementary excitations has the same form as the one in the quantum $^4He$ liquid (Landau – Bogolubov curve, Fig. 1).

Let us consider a non-ideal Fermi liquid with attraction between electrons. Landau developing a theory of quasi-particles in a non-ideal Fermi liquid supposed that the energy of the quasi-particles, being a functional of the distribution function, varies with that distribution function. That is

$$\delta\varepsilon = \int \Psi(\vec{p},\vec{p}',\vec{r})\, \delta f(\vec{r},\vec{p}',t) \frac{d^3p'}{(2\pi\hbar)^3}, \qquad (2)$$

where $\Psi(\vec{p},\vec{p}',\vec{r})$ is the quasi-particle interaction function. Obviously, in a Fermi ideal gas $\Psi = 0$.

In the case of attraction ($\Psi(\vec{p},\vec{p}',\vec{r}) < 0$) between electrons, the ground state of the system is unstable, which leads to the Cooper effect [9], i.e. to pair bound state formation of pair electrons lying near the Fermi surface in p-space and having equal and opposite momenta and antiparallel spins.

Near the surface of the Fermi sphere the variation of distribution function $\delta f(\vec{r},\vec{p}',t)$ is appreciably different from zero, i.e. the magnitude $p' = p = p_F$ ($p_F$ is the Fermi momentum). The same is true for the function $\Psi(\vec{p},\vec{p}',r) \approx \Psi(p_F,r)$. In such case the expression (2) can be rewritten as

$$\delta\varepsilon = \Psi(p_F,r)\delta n.$$

We now proceed with the derivation of the dispersion relation. To this end, we employ the quantum kinetic equations of the Fermi particles derived in Refs.[10,11], which read

$$\frac{\partial f_\alpha}{\partial t} + (\vec{v}\cdot\vec{\nabla})f_\alpha + e_\alpha\left(\vec{E} + \frac{\vec{v}\times\vec{H}}{c}\right)\frac{\partial f_\alpha}{\partial \vec{p}} + \vec{F}\frac{\partial f_\alpha}{\partial \vec{p}} + \frac{\hbar^2}{2m_\alpha}\vec{\nabla}\frac{1}{\sqrt{n_\alpha}}\Delta\sqrt{n_\alpha}\frac{\partial f_\alpha}{\partial \vec{p}} = 0, \quad \vec{F} = -\vec{\nabla}\,\delta\varepsilon, \qquad (3)$$

where suffix $\alpha$ stands for particle species.

Note that special and very important case in the quantum plasma is a one-fluid approximation. In this case one assumes that the characteristic dimension $R$ of inhomogeneities in the plasma is larger than the electron Thomas-Fermi length $r_{TF} = \frac{v_{Fe}}{\sqrt{3}\omega_{pe}}$. In the perturbed plasma the potential of the electric field is determined by Poisson equation

$$\Delta\varphi = 4\pi e(\delta n_e - Z_i \delta n_i). \qquad (4)$$

To estimate a magnitude of the term on the right-hand side of Eq. (4), we consider two cases $e\varphi \ll \varepsilon_F$ and $e\varphi \sim \varepsilon_F$. Noting that $\Delta\varphi \sim \varphi/R^2$ for the weak electric field, $\delta n_e = \frac{3}{2}n_0 \frac{e\varphi}{\varepsilon_F}$, we rewrite Eq. (4) as

$$\frac{\delta n_e - Z_i \delta n_i}{\delta n_e} = \frac{r_{TF}^2}{R^2} \ll 1. \qquad (5)$$

Remarkably, this inequality (5) is also satisfied for the strong perturbation, $e\varphi \sim \varepsilon_F$ and $\delta n_e \sim n_e$. Thus we can conclude that the uncompensated charge density is small compared to the perturbation of the electron and ion charge density separately. So, for $r_{TF} \ll R$ we further assume that the quasi-neutrality

$$n_e \simeq n_i \tag{6}$$

is satisfied. This equation along with the equation of motion of ions and the equation giving the adiabatic distribution of electrons allow us to define the potential field. In such approximation the charge is completely eliminated from the equations, and the Thomas-Fermi length $r_{TF}$ disappears with it.

In order to construct the one-fluid quantum kinetic equation, we neglect the time derivative in Eq.(3) of electrons, as well as the collision terms, assume $\vec{E} = -\nabla \varphi$ and $\vec{H} = 0$, and write the dynamic equation for the quasi-neutral plasma (6)

$$(\vec{v} \cdot \vec{\nabla}) f_e + \vec{\nabla} \left( e\varphi - \delta\varepsilon + \frac{\hbar^2}{2m_e} \frac{1}{\sqrt{n_e}} \Delta\sqrt{n_e} \right) \cdot \frac{\partial f_e}{\partial \vec{p}} = 0, \tag{7}$$

$$\frac{\partial f_i}{\partial t} + (\vec{v} \cdot \vec{\nabla}) f_i - \vec{\nabla} e\varphi \cdot \frac{\partial f_i}{\partial \vec{p}} = 0. \tag{8}$$

In Eq.(8) we have neglected the quantum term as a small one. Note that the Fermi distribution function of electrons

$$f_e = \frac{1}{exp\left\{\frac{\frac{p^2}{2m_e} - U - \mu_e}{T}\right\} + 1} \tag{9}$$

satisfies Eq.(7). Here $U = e\varphi - \delta\varepsilon + \frac{\hbar^2}{2m_e} \frac{\Delta\sqrt{n_e}}{\sqrt{n_e}}$ and $\mu_e$ is the chemical potential.

For the strongly degenerate electrons, i.e. $T_e \to 0$ ($\mu_e = \varepsilon_F$), the Fermi distribution function becomes the step function

$$f_e = \Theta\left(\varepsilon_F + U - \frac{p^2}{2m_e}\right), \tag{10}$$

which allows us to define the density of electrons $\left(n_e = n_i = n = 2\int \frac{d^3p}{(2\pi\hbar)^3} f_e\right)$

$$n = \frac{p_F^3}{3\pi^2 \hbar^3} \left(1 + \frac{e\varphi - \delta\varepsilon + \frac{\hbar^2}{2m_e} \frac{\Delta\sqrt{n}}{\sqrt{n}}}{\varepsilon_F}\right)^{3/2} \tag{11}$$

We now express $e\varphi$ from Eq.(11) and substitute it into the kinetic equation (8) to obtain

$$\frac{\partial f}{\partial t} + (\vec{v} \cdot \vec{\nabla}) f - \vec{\nabla} \left\{ \varepsilon_F \left(\frac{n}{n_0}\right)^{2/3} + \delta\varepsilon - \frac{\hbar^2}{2m_e} \frac{1}{\sqrt{n}} \Delta\sqrt{n} \right\} \frac{\partial f}{\partial \vec{p}} = 0. \tag{12}$$

This is the nonlinear kinetic equation of quantum plasma in the one-fluid approximation, which incorporates the potential energy due to the degeneracy of the plasma, the Madelung potential and the energy of quasi-particles.

We next derive a set of fluid equations. The Boltzmann and Vlasov type of quantum kinetic equation (12) gives a microscopic description of the way in which the state of the plasma varies with time. It is also well known how the kinetic equation can be converted into the usual equations of fluids. Following the standard method, one can derive the equations of continuity and motion of macroscopic quantities from equation (12)

$$\frac{\partial n}{\partial t} + \vec{\nabla}(n\vec{u}) = 0 \tag{13}$$

$$\frac{\partial \vec{u}}{\partial t} + (\vec{u} \cdot \vec{\nabla})\vec{u} = -\frac{T_{Fe}}{m_i} \vec{\nabla}\left(\frac{n}{n_0}\right)^{\frac{2}{3}} - \frac{1}{m_i} \vec{\nabla}\delta\varepsilon + \frac{\hbar^2}{2m_e m_i} \vec{\nabla} \frac{1}{\sqrt{n}} \Delta\sqrt{n} \tag{14}$$

where $\vec{u}(\vec{r}, t)$ is the macroscopic velocity of the plasma

$$\vec{u} = \frac{1}{n} \int \frac{2d^3p}{(2\pi\hbar)^3} \vec{v} f(\vec{r}, \vec{p}, t). \tag{15}$$

To consider the propagation of small perturbations $n = n_0 + \delta n(\vec{r}, t)$ and $\vec{u}(\vec{r}, t) = \delta\vec{u}(\vec{r}, t)$, we shall linearize equations (13) and (14) with respect to the perturbations, look for a plane-wave solution as $e^{i(\vec{k}\vec{r}-\omega t)}$, and derive the dispersion relation in the frequency range $kv_{Fe} \gg \omega \gg kv_{Fi}$, for the quantum Fermi liquid

$$\omega^2 = k^2 \left\{ \frac{p_F^2}{3m_e m_i} - \frac{U_0}{m_i}(1 + k\lambda_F) + \frac{\hbar^2 k^2}{4m_e m_i} \right\}. \tag{16}$$

Here noting $\delta\varepsilon(k,\omega) = \int d\vec{r} \int dt\, \delta\varepsilon(\vec{r},t) e^{-i(\vec{k}\vec{r}-\omega t)}$, use was made of

$$\delta\varepsilon(k,\omega) = -U_0(1 + k\lambda_F)\frac{\delta n(k,\omega)}{n_0}, \tag{17}$$

where $U_0 = |\Psi(p_F)|n_0$, $\lambda_F = \frac{\hbar}{p_F}$ and $k\lambda_F < 1$.

We now show from the dispersion relation (16) that for the Fermi liquid exists the same type of curve as the one for the Bose liquid $^4He$ (Fig.1). To this end, we differentiate the dispersion equation (16) by $k$ to obtain

$$\frac{d\omega}{dk} = \frac{k}{2m_i\omega}\left\{ \frac{\hbar^2 k^2}{m_e} - 3U_0\lambda_F k + \frac{2}{3}\left(\frac{p_F^2}{m_e} - 3U_0\right)\right\}. \tag{18}$$

From which we find the points of extreme, $\frac{d\omega}{dk} = 0$,

$$k_1 = \frac{3}{2}\frac{m_e U_0}{\hbar p_F}\left\{ 1 - \sqrt{1 - \frac{8}{3}\left(\frac{p_F^2}{3m_e U_0}\right)^2\left(1 - \frac{3m_e U_0}{p_F^2}\right)} \right\} \tag{19}$$

and

$$k_2 = \frac{3}{2}\frac{m_e U_0}{\hbar p_F}\left\{ 1 + \sqrt{1 - \frac{8}{3}\left(\frac{p_F^2}{3m_e U_0}\right)^2\left(1 - \frac{3m_e U_0}{p_F^2}\right)} \right\}. \tag{20}$$

Obviously at

$$\frac{8}{3}\left(\frac{p_F^2}{3m_e U_0}\right)^2\left(1 - \frac{3m_e U_0}{p_F^2}\right) < 1 \tag{21}$$

both $k_1$ and $k_2$ are real ($k_1 < k_2$). Moreover, it is clear from Eq.(21) that the following relation should hold, $3m_e U_0 \sim p_F^2$.

It should be emphasized that solutions (19) and (20) with the condition (21) describe the similar curve as the Landau-Bogolubov's (Fig.1) that was obtained for the quantum $^4He$ liquid. Namely, if $k < k_1$ and $k > k_2$, then the derivative $\frac{d\omega}{dk}$ is positive, whereas for $k_1 < k < k_2$ is negative and in such case we get for $\omega(k)$ the same form as in $^4He$ liquid (Fig. 1), i.e. the energy spectra of elementary excitations for superfluid $^4He$ liquid and superconductor are the same type.

To summarize, we have derived the dispersion relation for the quantum Fermi liquid and established that the energy spectra in the Fermi liquid (attraction interaction) and the Bose liquid $^4He$ (repulsion interaction), when they are superfluid states, have identical curves (Fig. 1). Thus, we have shown that the superconductivity is, in fact, the same as superfluidity, but for charged particles. We should also note that the kinetic energy of electrons in metal, which is of the order of the Fermi energy $\varepsilon_F \sim 1 - 10\ eV$, is not at all high compared to the energy of quasi-particles $\varepsilon_F \sim U_0 \sim |\Psi|n_0$. The theory developed in the present paper is of great general theoretical interest, as well as may be of substantial practical use, for understanding of a broad range of materials and contemporary problems in superconductivity.